\documentclass[fleqn,10pt]{wlscirep}
\usepackage[utf8]{inputenc}
\usepackage[T1]{fontenc}
\usepackage{graphicx}
\usepackage{subcaption}    
\usepackage{float}
\usepackage{amsmath}
\usepackage{upgreek}
\usepackage{amssymb}
\usepackage{setspace}
\usepackage{indentfirst}
\usepackage{overpic}       
\usepackage{hyperref}
\usepackage{diagbox}

\title{Agency Perception and Brain Synchrony: A Hyperscanning Study of Human–Human and Human–AI Interaction}

\author[1,*]{Mohammad Ghalavand}
\author[1, +]{Javad Hatami}
\author[2, +]{Seyed Kamaledin Setarehdan}
\author[1, +]{Fatimah Nosrati}
\author[3, +]{Hananeh Ghalavand}
\author[4]{Ali Nikhalat-Jahromi}
\affil[1]{Department of Psychology and Educational Sciences, University of Tehran, Tehran, Iran}
\affil[2]{School of Electrical and Computer Engineering, College of Engineering, University of Tehran, Tehran, Iran}
\affil[3]{School of Pharmacy, Shahid Beheshti University of Medical Sciences, Tehran, Iran}
\affil[4]{Department of Computer Engineering, Amirkabir University of Technology, Tehran, Iran}

\affil[*]{mohammadghalavand@ut.ac.ir}
\affil[+]{these authors contributed equally to this work}

\keywords{Theory of Mind, Artificial Intelligence, fNIRS, Prefrontal Cortex, Intentional Agent}

\begin{abstract}
This study investigates how the human brain differentiates between intentional human agents and artificial intelligence (AI) agents during real-time social interaction. Using functional near-infrared spectroscopy (fNIRS) hyperscanning, we recorded prefrontal brain activity of participants as they played a one-on-one virtual tennis game, once against a human opponent and once against an AI-controlled opponent. Task-related cortical activation was evaluated using the General Linear Model (GLM), while inter-brain synchrony was assessed through Intersubject Correlation (ISC) analysis. Results revealed significantly stronger activation in the medial prefrontal cortex (mPFC) when participants interacted with human opponents, particularly in low-frequency hemodynamic bands (0.0098–0.0196 Hz). Moreover, neural synchrony between human dyads was significantly greater than in pseudo-paired controls, with distinct frequency-specific coupling patterns across channels. In contrast, AI interactions elicited reduced cortical engagement and no measurable inter-brain synchrony. These findings suggest that the perception of intentionality strongly modulates social brain responses, even when task structure and behavior remain constant. This work highlights the utility of fNIRS hyperscanning for studying naturalistic human–agent interactions and provides neural evidence for the brain’s sensitivity to perceived agency. The results have implications for the design of socially interactive AI systems that more effectively engage human cognition.
\end{abstract}

\begin{document}

\flushbottom
\maketitle
%
%
\thispagestyle{empty}

\section*{Introduction}

Human beings are inherently social, equipped with sophisticated cognitive abilities that have evolved to navigate interpersonal interactions. A central component of human social cognition is the capacity for \textbf{theory of mind} (ToM) – the ability to attribute thoughts, intentions, and emotions to others\cite{bermudez2020cognitive,fiske2013social}. This mentalizing capacity allows individuals to interpret and predict the behavior of people around them, forming the foundation of communication, empathy, and cooperation. In interactive settings, theory of mind is continually engaged as we read subtle social cues, anticipate others’ actions, and adjust our own behavior accordingly. Underlying these skills is a dedicated social brain network encompassing regions such as the \textit{prefrontal cortex} and \textit{temporoparietal junction}, which become active when we consider another’s perspective or engage in real-time social exchange \cite{rilling2004neural, saxe2006neuroscience}.

Traditionally, neuroscience research on social interaction has been conducted under controlled laboratory conditions, often focusing on isolated individuals responding to static stimuli. While these studies have been invaluable for identifying key brain areas involved in social cognition, they capture only a slice of the rich, dynamic nature of true interpersonal engagement. Real-world interactions are fluid and reciprocal – people exchange signals back and forth, whether through conversation, cooperative tasks, or competitive games. In such interactive scenarios, both participants influence each other’s behavior and brain activity in real time\cite{babiloni2014social}. This has led to growing recognition that to truly understand social cognitive processes, scientific investigations must move beyond one-sided tasks and embrace \textbf{two-person paradigms} that capture the bidirectional flow of communication. Indeed, emerging research using hyperscanning\cite{czeszumski2020hyperscanning} (simultaneous neuroimaging of two interacting people) highlights that brains engaged in direct interaction exhibit patterns that differ from those observed in solo settings \cite{carollo2024hyperscanning,hamilton2021hyperscanning}.

Compounding this need is the changing landscape of social interaction in modern society. With rapid advances in technology, humans are increasingly interacting not only with other humans but also with \textbf{artificial intelligence (AI)} systems on a daily basis\cite{tegmark2017life}. From virtual assistants and chatbots to more advanced AI-driven avatars and robots, these artificial agents are beginning to occupy roles as collaborators\cite{chan2020artificial}, opponents, or companions\cite{fabre2024making} in various domains \cite{vantrepotte2022leveraging,salatino2025influence}. This raises important questions about social cognition: Do our brains respond to an AI partner in the same way as to a human partner?

A human being represents an \textit{intentional agent} – an entity presumed to have its own independent mind \cite{wiese2017robots,perez2020adopting}, desires, and goals. In contrast, even the most sophisticated AI is fundamentally a \textit{non-intentional agent}, guided by algorithms without true human-like intentions or consciousness. Intuitively, one might expect that interacting with a fellow human would engage deeper instinctual social processes (like empathy or mentalizing) than interacting with a machine. However, AI systems are becoming ever more human-like in their behavior, blurring the lines and perhaps prompting humans to subconsciously treat them in a social manner \cite{chan2020artificial,wiese2017robots}.

In this context, understanding how the brain differentiates (or fails to differentiate) between intentional and artificial agents is both scientifically intriguing and practically relevant. It speaks to the core of how flexible our social cognition is and how it might adapt in an age where human–AI interactions are commonplace. Prior studies in social neuroscience have begun to shed light on this issue by comparing human brain responses during interactions with people versus computers. Results from various experimental games and decision-making tasks indicate that the presence of another human mind can alter neural activation patterns. For example, when individuals believe they are competing or cooperating with a real person (as opposed to a computer program), they often show greater engagement of brain regions associated with theory of mind and social evaluation, such as the \textit{medial prefrontal cortex (mPFC)} \cite{gallagher2002imaging,krach2010rewarding}.

This suggests that humans adopt an \textbf{intentional stance} toward other people – automatically considering the invisible mental states behind a partner’s actions – whereas they may adopt a more mechanical or analytical stance toward a machine opponent \cite{dennett1988precis}. Even in relatively simple paradigms (like turn-based games or economic exchanges), participants tend to exhibit heightened activation in the social brain network and distinct emotional responses if they perceive their counterpart as another person. These findings underscore that the cognitive and neural processing during an interaction is shaped by whom (or what) we interact with. However, much of this evidence comes from tightly controlled settings that strip down the interaction to basic elements. The challenge and next step are to examine these differences in brain responses within naturalistic, real-time interactions, which better represent how we engage with others in daily life.

In pursuit of that realism, the present study employs \textbf{functional near-infrared spectroscopy (fNIRS)} to investigate brain activity during interactive gameplay. fNIRS is a neuroimaging technique well-suited for studying live social interactions: it uses light sensors placed on the scalp to monitor changes in blood oxygenation in the cortex, providing an index of neural activity \cite{cui2011quantitative,pinti2020present,pinti2019current}. Unlike fMRI or PET scanning, which require participants to lie still in a scanner, fNIRS devices are portable and tolerant of movement, allowing participants to sit in a normal environment or even interact face-to-face. This makes it feasible to bring neuroscience out of the lab and into more dynamic settings.

To create a realistic yet controlled interactive scenario, we designed a \textbf{virtual tennis game} as the platform for social engagement in this study. A tennis match is an inherently interactive and dynamic activity: players must continuously anticipate their opponent’s moves, formulate strategies, and respond within fractions of a second. These demands closely mirror everyday social cognitive challenges, where one must infer others’ intentions and react adaptively. In our experimental setup, participants play a one-on-one tennis simulator, which provides a standardized environment with consistent rules and feedback. This game-based paradigm offers several advantages for studying real-time social cognition.

A key feature of our study is the comparison between two modes of interaction within this tennis game: \textbf{playing against a human opponent} versus \textbf{playing against an AI opponent}. In the human-opponent condition, the participant believes (and genuinely has) a real person on the other side of the game, with whom they are competing in real time. This scenario embodies an interaction with an intentional agent, likely engaging the full spectrum of social cognitive processing. In the AI-opponent condition, the participant faces a computer-controlled player governed by artificial intelligence algorithms. This represents an interaction with a non-human agent. By confronting the same individual with these two contrasting opponent types, our study can isolate how the brain’s response depends on the perceived agency and intentionality of the opponent.

Beyond examining regional brain activation, our study also acknowledges the interactive nature of human-human play by considering \textbf{neural synchrony} between players. When two people engage directly, especially in cooperative or adversarial settings, their neural activity can become temporally aligned in certain frequency bands or regions – a phenomenon known as \textit{interbrain synchrony}\cite{miller2019inter,reveille2024using}. In a competitive game like tennis, such coupling would be a hallmark of genuine social interaction that has no equivalent when one plays against a machine. While an AI can mimic behavior, it does not share a brain, and thus any rhythm or coordination arises from the human’s adaptation to an algorithm rather than true mutual alignment.

\textbf{In summary}, this study aims to explore whether interactions with AI engage the social brain in a manner comparable to interactions with humans, or whether they elicit fundamentally different patterns of neural activity. Using the ecological flexibility of fNIRS and a dynamic game-based paradigm, we assess neural responses during one-on-one gameplay against human and artificial opponents. The outcomes are expected to inform theoretical models of social cognition and contribute to the design of AI systems that better align with human social processing.

\section*{Materials and Methods}

\subsection*{Ethical Approval}
This study was approved by the Research Ethics Committee of the Faculty of Psychology and Education, University of Tehran, under the code IR.UT.PSYEDU.REC.1401.074. The committee thoroughly reviewed the research protocol and confirmed that all procedures complied with the highest ethical standards. Moreover, all procedures were conducted in accordance with the guidelines and regulations of the Iranian National Brain Mapping Lab Ethics Committee. Written informed consent was obtained from all participants prior to participation, in line with the approved ethical framework.

\subsection*{Participants}
A total of 50 right-handed male participants (mean age = 24.94 years, SD = 3.77; age range: 18–30) took part in the study. Participants were screened to exclude those with a history of psychiatric, neurological, or medical conditions, as well as those who had taken psycho-pharmacological medication within the past two months. Eligibility criteria also required participants to have basic knowledge of tennis, previous experience with computer-based games, and no prior exposure to professional tennis simulators. These criteria were implemented to minimize variability in baseline skill and ensure consistency in task performance.

The decision to include only male participants was based on both methodological and practical considerations. A substantial body of literature in cognitive and social neuroscience suggests that men and women may engage distinct neural pathways and strategies during social information processing tasks \cite{hall2012social,pearce2019genetic,proverbio2023sex}. Additionally, sex-related differences in brain morphology and functional modularity have been reported, which could introduce additional variability in fNIRS signal interpretation \cite{baker2016sex,shirzadi2024sex}. Given that fNIRS has limited spatial resolution compared to other neuroimaging modalities (e.g., fMRI), and since precise anatomical mapping often relies on MRI co-registration—which was beyond the scope of the present study—we opted to reduce variance by restricting the sample to a single gender.

Furthermore, recruitment feasibility influenced this decision: our advertisements specified a minimum familiarity with computer games, which resulted in a higher number of eligible male respondents. Taken together, these factors motivated the use of an all-male sample in this exploratory study.

\subsection*{fNIRS Data Acquisition}
Functional near-infrared spectroscopy (fNIRS) data were recorded using a 6-channel OxyMon system (Artinis Medical Systems, The Netherlands), which operates at wavelengths of 740 nm and 860 nm. These wavelengths enable spectroscopic differentiation of oxygenated and deoxygenated hemoglobin. The data were sampled at 10 Hz to capture real-time hemodynamic responses during gameplay.

Optode placement followed the international 10–20 system \cite{homan198810}, with a primary focus on the medial prefrontal cortex near FPz—approximately corresponding to Brodmann areas 9 and 10 \cite{okamoto2004three}—which are known to be engaged during mentalizing and theory of mind processes \cite{pinti2021role}. Four additional channels targeted lateral prefrontal regions (F7 and F8), corresponding to Brodmann areas 45–47 \cite{okamoto2004three, herwig2003using}, which are implicated in inhibitory control and altruistic behavior \cite{himichi2015modulation, lague2013fnirs}.

Although other brain areas such as the temporoparietal junction (TPJ) have been strongly associated with social cognition in recent literature, our study prioritized the prefrontal cortex for several reasons. First, we aimed to preserve participants' natural movement and interactivity as much as possible during the game. Based on our pilot tests, the prefrontal cortex exhibited relatively high signal-to-noise ratio and robustness to motion artifacts, making it an optimal target for mobile neuroimaging. Second, precise spatial localization of socially relevant regions such as the TPJ has been extensively achieved in previous studies using high-resolution imaging techniques such as fMRI and PET \cite{gallagher2002imaging,kramer2010emotional,van2009social}. Our main goal here was not to replicate those precise mappings, but rather to extend such findings into more ecologically valid settings using a mobile and participant-friendly modality such as fNIRS.

\begin{figure}[H]
    \centering
    
    \begin{subfigure}[b]{0.54\textwidth}
        \includegraphics[width=\textwidth]{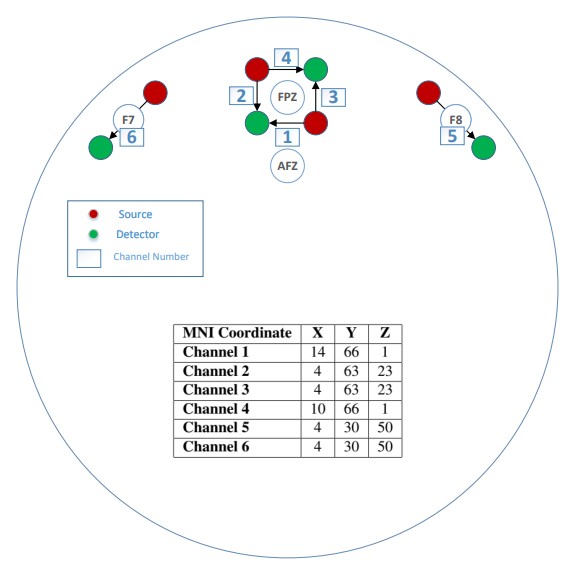}
        \caption{}
       
    \end{subfigure}
    \hfill
    \begin{subfigure}[b]{0.44\textwidth}
        \includegraphics[width=\textwidth]{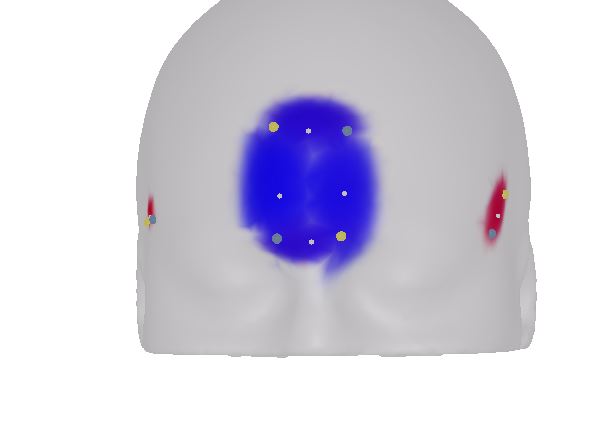}
        \caption{}
        
    \end{subfigure}
    \caption{a: Optode placement according to the 10–20 system and corresponding MNI coordinates. b: The image is provided by the fNIRS device from Artinis, and it approximately shows the cortical areas involved in the frontal and prefrontal regions, as well as the lateral parts.}
    \label{fig:channel_position}
\end{figure}

Given this targeted placement, it was important to accurately register optode locations in standard brain space to allow for anatomical interpretation and potential comparison with previous neuroimaging studies. To project optode positions into MNI space, we employed the fOLD toolbox \cite{zimeo2018fnirs} within the MATLAB environment. The resulting standardized channel coordinates are illustrated in Figure~\ref{fig:channel_position}.

\subsection*{AO Tennis 2 Game as a Mind Reading Task}
In this study, a tennis simulator computer game was employed, which proved to be well-suited to this research objectives for several reasons. This game offered several advantages, such as ease of learning for the participants, the ability to be played by both single and two players, and the adaptability of game difficulty levels through simple adjustments, ensuring the engagement of the participants. 

Unlike the static nature of Rock-Paper-Scissors that is used in similar works, tennis exemplifies dynamic gameplay with complete information, mirroring real-life situations. Players make sequential decisions, reacting to their opponent's moves in real-time, and possess full awareness of the game state. This parallels many real-life scenarios where individuals must adapt and make strategic decisions based on complete information to achieve success \cite{tadelis2013game, shoham2008multiagent}.

\begin{figure}[H]
\centering
\includegraphics[width=1\textwidth]{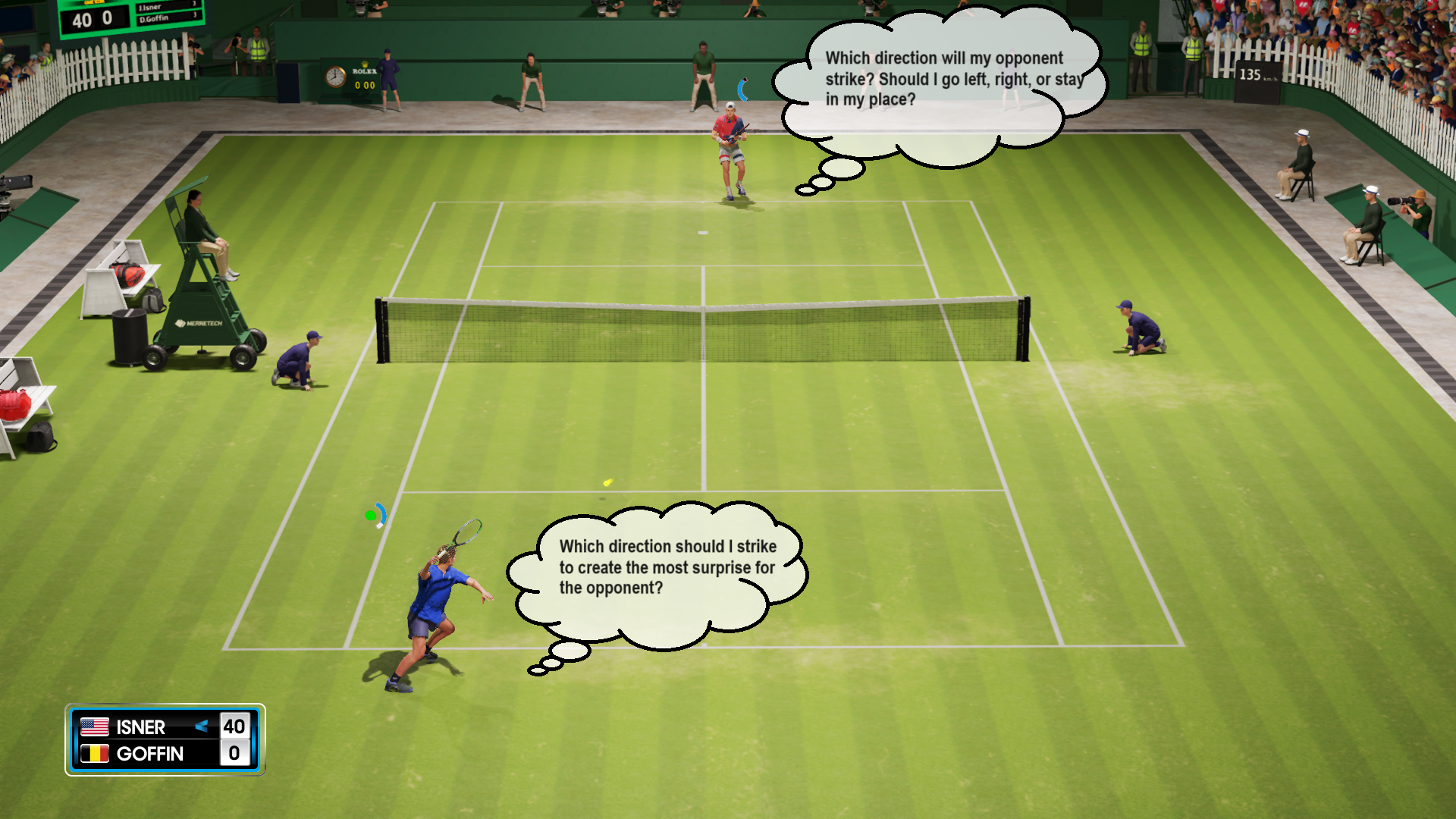}
\caption{The game environment. Participants must constantly anticipate the direction of their opponent's strike. While striving to reach the ball, each competitor must calculate which direction holds more surprises for their opponent.}
\label{fig:ping_pong}
\end{figure}

This game demonstrated excellent flexibility for implementing a block design, a key feature for experimental setup. What truly sets this game apart, however, is its alignment with the nature of tennis itself. Tennis is a dynamic sport, demanding simultaneous planning for the next shot while strategizing to counter the opponent's moves. This intricate process necessitates the continuous engagement of cognitive modules related to theory of mind and complex calculations within the brain. The game environment is shown in Fig.\ref{fig:ping_pong}.

\subsection*{Experimental Paradigm}
\begin{figure}[H]
\includegraphics[width=0.9\textwidth]{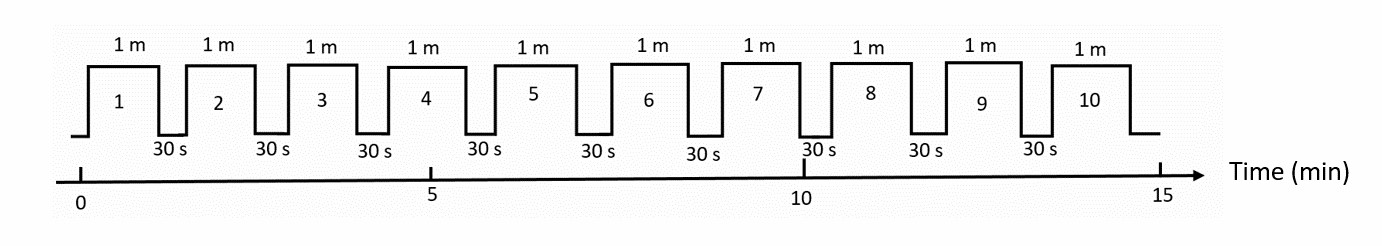}
\caption{Each game comprises ten cycles, with each cycle consisting of a 1-minute gameplay session followed by a 30-second rest period.}
\label{fig:block_design}
\end{figure}

\begin{figure}[H]
    \centering

    \begin{subfigure}[b]{0.45\textwidth}
        \includegraphics[width=\textwidth, height=0.45\textheight]{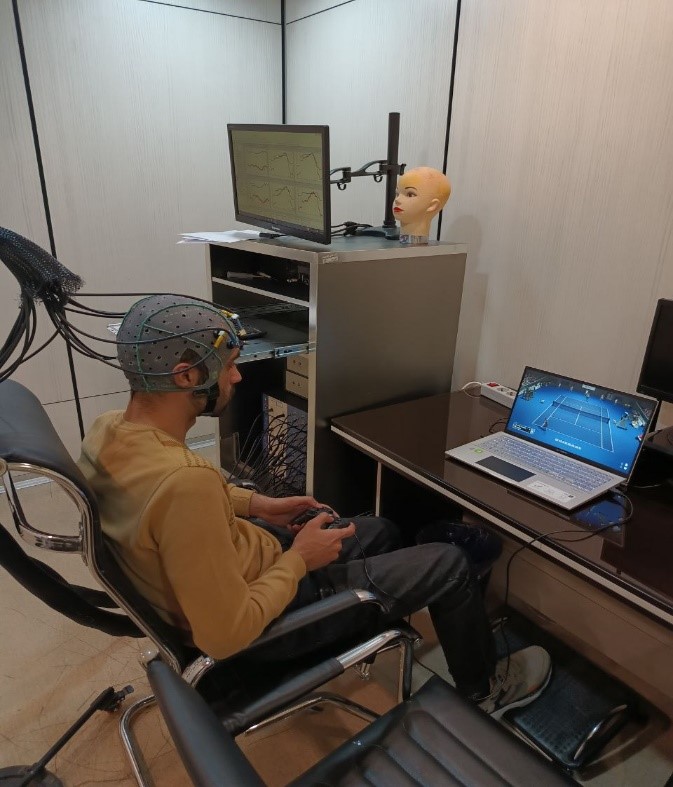}
        \caption{}
        \label{fig:env1}
    \end{subfigure}
    \hfill
    \begin{subfigure}[b]{0.45\textwidth}
        \includegraphics[width=\textwidth, height=0.45\textheight]{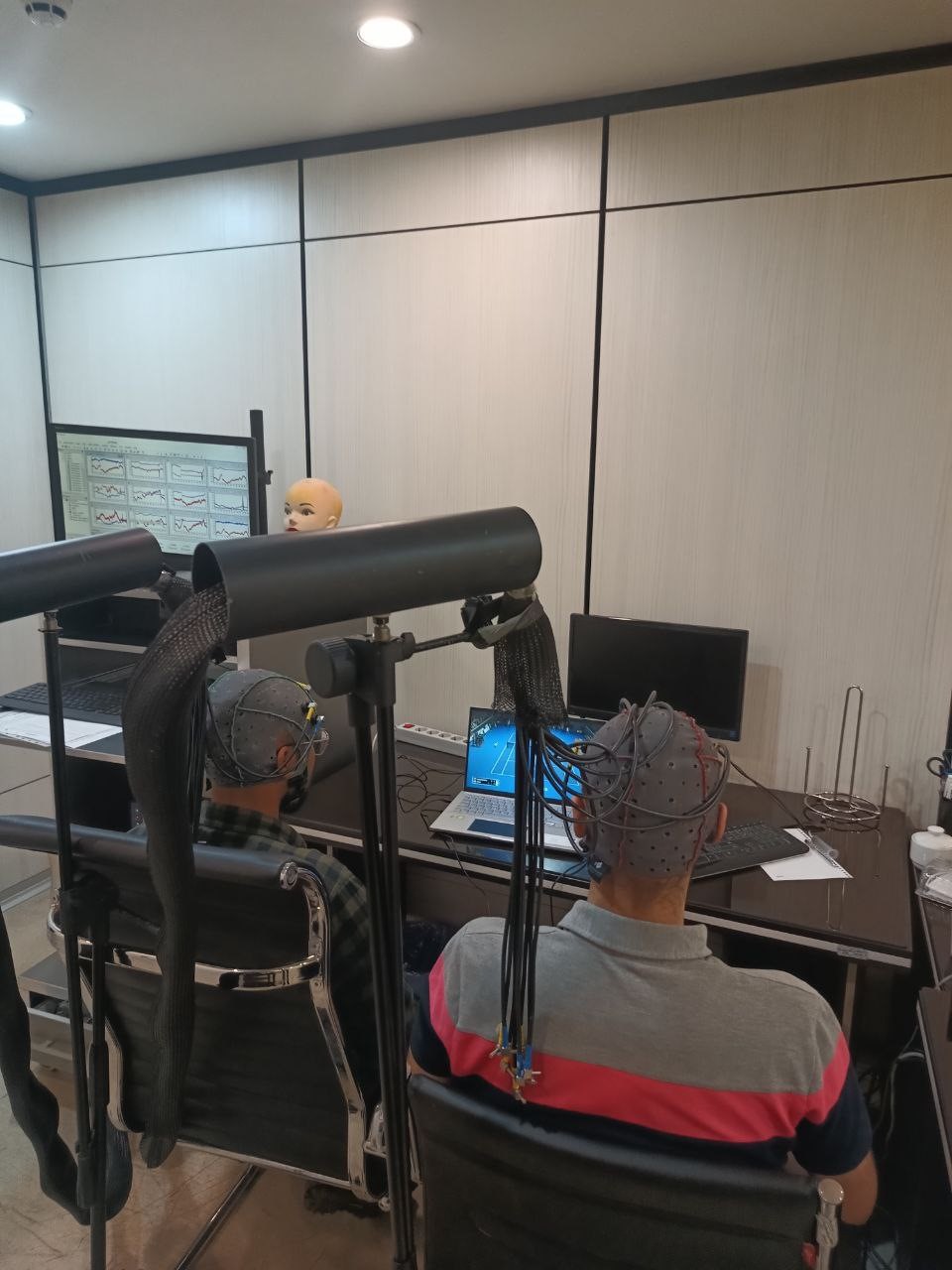}
        \caption{}
        \label{fig:env2}
    \end{subfigure}

    \vspace{1em} 

    \begin{subfigure}[b]{0.3\textwidth}
        \centering
        \includegraphics[width=\textwidth]{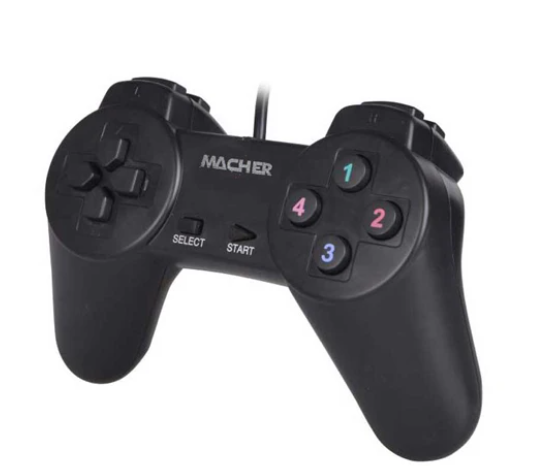}
        \caption{}
        \label{fig:joystick}
    \end{subfigure}

    \caption{
        Scenes from the laboratory environment and the participants engaging in the game. 
        Image (a) depicts a single-player game with AI, 
        image (b) shows a game with a human counterpart, 
        and image (c) shows the joystick or controller used for the game. It typically involved three buttons: left, right, and button number 2 (used for a simple hit action). This simplicity was intended to ensure all participants were on an equal footing.
    }
    \label{fig:imgs}
\end{figure}

During the experimental phase, participants engaged in two 15-minute gaming sessions, involving a simulated tennis video game (Fig.\ref{fig:imgs}), while simultaneously undergoing fNIRS scanning (Fig.\ref{fig:block_design}). The experimental design encompassed two distinct scenarios: solo gameplay against an AI opponent and two-player gameplay against a human opponent, all of which were monitored through fNIRS hyperscanning.

To ensure precise timing and synchronization, the experimenter issued verbal commands. Participants were prompted to "stop" and close their eyes, and upon hearing "continue," they resumed gameplay. A timer was meticulously programmed to maintain accurate timing, and the experimenter used a remote controller to pause and resume the game at the initiation and conclusion of each gaming cycle.

In the solo gameplay scenario, each participant, referred to as Participant 1 and Participant 2, individually completed a gaming session against a computer opponent. This gameplay involved a series of cycles, with each participant engaging in tennis matches against the AI for 1 minute, followed by 30 seconds of rest. This cycle was repeated ten times, resulting in a total duration of 15 minutes.

In the two-player scenario, Participant 1 and Participant 2 engaged in a live tennis match against each other, with simultaneous fNIRS hyperscanning. This two-player session mirrored the solo condition, maintaining the structured alternation between 1-minute gameplay and 30-second rest cycles, also totaling 15 minutes. Participants were explicitly instructed to strive for victory in both the computer and human opponent scenarios. Incentives were provided in the form of performance-based scores, in addition to their base participation compensation. In the two-player human condition, the final score directly reflected the game's outcome. In contrast, in the solo AI condition, the score was based on surpassing the AI opponent's score, creating an indirect competition. To maintain equal effort in both gaming scenarios, participants remained unaware of the final score's value, which was randomly determined through a coin flip after the sessions, ensuring their complete engagement in both human and AI gameplay. 

\section*{Data Analysis}
Data analysis played a pivotal role in unraveling the complex neural dynamics underlying this study. The fNIRS data underwent a rigorous and comprehensive preprocessing phase, which aimed not only to enhance data quality but also to eliminate potential artifacts that could affect the validity of the findings. This section provides a detailed account of the steps taken during data analysis, from preprocessing to statistical analysis.

\begin{figure}[H]
    \centering
    \includegraphics[width=0.95\textwidth]{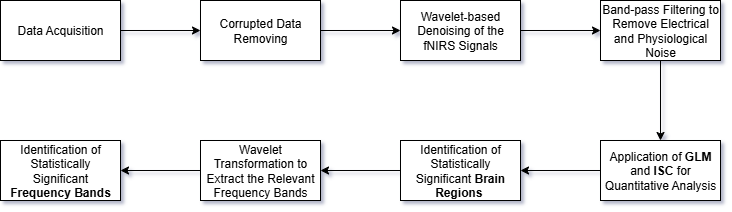}
    \caption{Overview of the fNIRS signal processing and analysis pipeline: starting from data acquisition and removal of corrupted signals, followed by wavelet-based denoising and band-pass filtering to eliminate noise. Quantitative analysis is then performed using GLM and ISC, leading to the identification of statistically significant brain regions and frequency bands through wavelet transformation}
    \label{fig:data_analysis}
\end{figure}

It is crucial to emphasize that this study's findings rely on a two-dimensional approach facilitated by the wavelet method. In this framework, one dimension corresponds to the chosen channel, while the other dimension corresponds to a specific frequency range.Fig.\ref{fig:data_analysis}.

\subsection*{Utilizing OxyHb and DeoxyHb Data for Enhanced fNIRS Analysis}
In analyzing our fNIRS data, it's essential to recognize the application of the Beer-Lambert law \cite{ferrari2004principles, delpy1997quantification}, which correlates the absorption of light in a medium with the concentration of absorbing substances such as hemoglobin. The Beer-Lambert law provides a fundamental basis for quantifying changes in hemoglobin concentration, crucial for interpreting neural activity measured by fNIRS. Our study utilized a device that has already applied the Beer-Lambert law to the raw data, yielding concentrations of both oxygenated (oxyHb) and deoxygenated (deoxyHb) hemoglobin.

While some previous analyses focused solely on oxyHb data \cite{hoshi2016hemodynamic, scholkmann2014review, yucel2021best, kirilina2012physiological, balconi2019donate, costantini2013studying}, we have expanded our investigation to include deoxyHb information as well. This decision aligns with standardized procedures highlighted in Yücel et al.'s paper on best practices for fNIRS publications \cite{yucel2021best}. Incorporating both oxyHb and deoxyHb data is crucial for providing a comprehensive understanding of neural activity, particularly in scenarios where physiological interferences related to the task may occur. For instance, previous research has indicated that individuals may alter their breathing patterns during tasks such as video gaming \cite{tachtsidis2013investigation}. By examining both oxyHb and deoxyHb responses, we aim to capture the complete hemodynamic response and provide a more nuanced interpretation of neural activity in response to our experimental conditions \cite{pinti2020present, costantini2013studying}.



\subsection*{Corrupted Data Removing}
Out of the initial 50 data samples collected using fNIRS technology, a total of 5 samples were excluded from further analysis due to various issues. One dyadic sessions were excluded due to large portions of missing signal data. One session was omitted because a participant withdrew mid-way through the game. Additionally, one dataset was excluded due to non-removable artifacts. As a result, 45 high-quality data samples were retained for subsequent mathematical and statistical analyses.

\subsection*{band pass and  Wavelet-Based Approach to Decompose Signal}

Initially, we used a 4th-order bandpass Butterworth filter to restrict the data to our desired frequency range of 0.01 to 0.09 Hz, which is relevant for cognitive tasks. This filtered data was then used for subsequent analyses. Although the primary method of analysis was bandpass filtering, we later employed wavelet analysis as a complementary approach to deepen our insights.

To enhance measurement accuracy and introduce additional depth to the data analysis, we applied the Daubechies wavelet at 10 decomposition levels to the signal. This segmentation resulted in the division of the data into 10 distinct components. The wavelet-based approach offers several advantages in processing fNIRS data, particularly in capturing both temporal and frequency-domain information simultaneously.

Previous fNIRS studies have shown that the frequency band of 0.01 to 0.1 Hz is typically associated with cognitive processes such as attention, prediction, and mind-reading, while spontaneous hemodynamic oscillations are related to physiological vasomotor regulation and breathing-related fluctuations. These are characterized by signals at specific frequencies, including heart rate (~1 Hz), breathing rate (~0.3 Hz), Mayer waves (~0.1 Hz), and very low-frequency oscillations (<0.04 Hz). Therefore, our interest lay in analyzing four specific frequency bands: level 6 (0.0779–0.157 Hz), level 7 (0.039–0.0874 Hz), level 8 (0.0195–0.0392 Hz), and level 9 (0.00978–0.0196 Hz), all of which fall within our target cognitive range.

While previous studies typically did not limit the frequency range to this extent and mostly relied on band-pass filters, cognitive-related effects have consistently been observed within this frequency band. The wavelet-based multiresolution analysis thus served as a complementary tool to further refine our understanding of the signal components.

\begin{figure}[!t]
\centering
\includegraphics[width=1\textwidth]{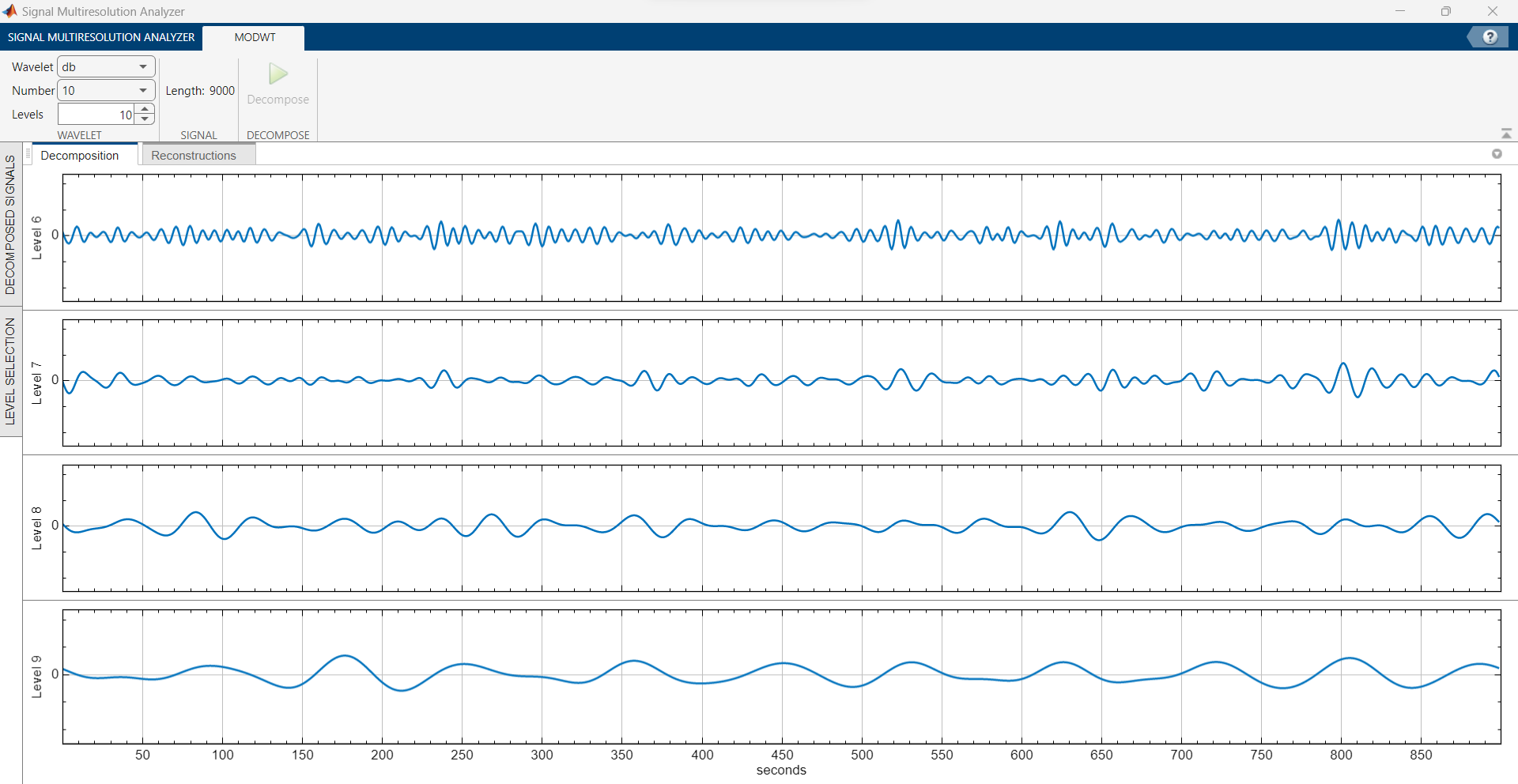}
\caption{Signal decomposition with Daubechies Wavelet of type and level 10 to 4 levels illustrates levels \textbf{6} (0.0779-0.157 Hz), \textbf{7} (0.039-0.0874 Hz), \textbf{8} (0.0195-0.0392 Hz), and \textbf{9} (0.00978-0.0196 Hz) for frequency intervals.}
\label{fig:decompose}
\end{figure}

\subsection*{Quantifying Brain Activity: GLM and ISC Approaches}

To comprehensively assess neural dynamics during gameplay, we employed two complementary analytical strategies: the General Linear Model (GLM) to examine localized task-evoked activation, and Intersubject Correlation (ISC) to assess inter-brain synchrony.

\begin{figure}[!t]
\centering
\includegraphics[width=0.7\textwidth]{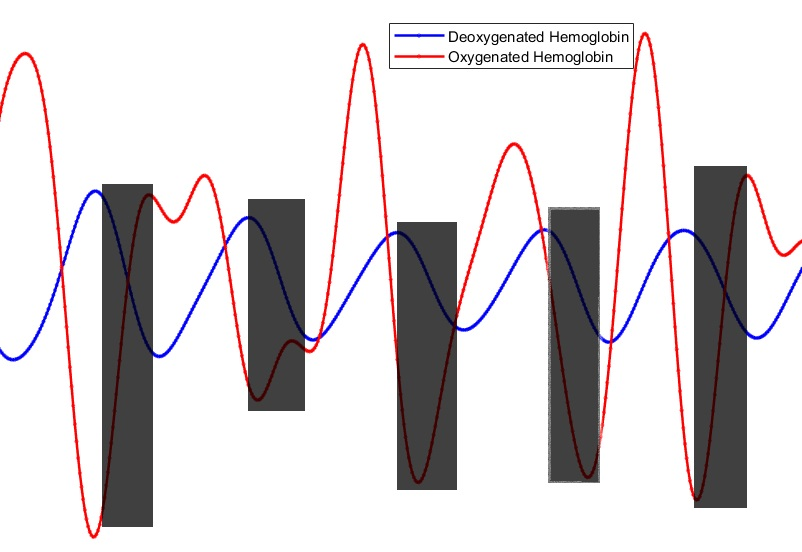}
\caption{Comparison of HbO and HHb signals. The GLM model imposed expected activity (1) during gameplay and inactivity (0) during rest.}
\label{fig:hemoglobin}
\end{figure}

\textbf{GLM-Based Activation Analysis.} For each participant, we modeled the fNIRS time series using a block design comprising alternating 60-second task and 30-second rest periods, repeated over a 15-minute session. A binary regressor (1 for task, 0 for rest) captured the experimental structure (see Fig.~\ref{fig:hemoglobin}). Using this design matrix, we estimated beta coefficients for each channel to quantify task-related activation. These coefficients were then averaged across participants to evaluate condition-specific cortical engagement.

\textbf{ISC-Based Synchrony Analysis.} To investigate the degree of neural coupling between participants, we computed ISC within non-overlapping 90-second windows (900 samples at 10 Hz), separately for each channel. ISCs were averaged across blocks, yielding a mean ISC per pair per channel. True interacting dyads (in the human vs. human condition) were contrasted with pseudo-dyads constructed from independent individuals in the single-player condition, providing a control distribution for synchrony arising by chance.

\subsection*{Nonparametric Statistical Inference for Activation and Synchrony}

We employed a unified nonparametric framework to statistically evaluate differences in both task-evoked activation (GLM)\cite{von2020improved,pinti2017novel} and inter-brain synchrony (ISC).

\textbf{GLM Inference.} Beta coefficients for each channel (separately for HbO and HbR) were compared within subjects (human-agent vs. AI-agent) using the Wilcoxon signed-rank test. To correct for multiple comparisons across six channels, we applied Bonferroni adjustment ($\alpha = 0.05/6$). Effect sizes (Cohen's $d$) and post-hoc statistical power were also calculated to assess the magnitude and reliability of the observed effects.

\textbf{ISC Inference.} For ISC, we compared true interacting pairs to a large pool of pseudo-pairs using the Mann–Whitney U test. Again, Bonferroni correction was applied across channels, and we report Cohen’s $d$ and power estimates. This nonparametric approach accounts for the unequal group sizes and skewed ISC distributions commonly found in hyperscanning data.

\textbf{Combined Interpretation.} Together, these analyses provide a dual perspective: GLM captures individual-level cortical responses, while ISC reflects interpersonal neural alignment. By incorporating corrected significance levels, effect size metrics, and statistical power, our framework ensures robust and interpretable inference on both intra- and inter-brain processes.

\subsection*{Exploratory Machine Learning Analysis to Address Order Effects}
To examine whether the fixed sequence of gameplay conditions (single-player always preceding two-player) may have introduced systematic temporal patterns unrelated to agent intentionality, we conducted a supplementary analysis using a Long Short-Term Memory (LSTM)\cite{shiri2023comprehensive} neural network. This approach aimed to identify whether neural signals contained sufficient temporal features to discriminate between single- and two-player sessions—potentially reflecting fatigue, skill acquisition, or increasing familiarity with the game mechanics.

The LSTM model comprised two recurrent layers with 128 and 64 units, each followed by dropout layers (with a rate of 0.3), and a fully connected dense layer with 32 ReLU-activated units. A final softmax layer was used for binary classification. The model was trained with the Adam optimizer and categorical cross-entropy loss function. Early stopping was employed to prevent overfitting, using a validation-based criterion with a patience of 10 epochs.

The input to the model consisted of 60-second time-course segments extracted from the fNIRS recordings. For each participant, we obtained 10 one-minute signals per condition, resulting in 450 samples for the single-player mode and 450 for the two-player mode. These windowed time-series were used as independent training instances.

If the model fails to achieve meaningful classification accuracy, this suggests the absence of a consistent trend across participants—indicating that confounding effects such as fatigue or learning curves did not systematically bias the brain responses across conditions. The detailed results of this control analysis are presented in the Results section.

\section*{Results}

To assess our hypothesis that distinct prefrontal channels support task‐evoked hemodynamic activation and inter‐brain synchrony, we conducted two parallel analyses.  First, we applied a General Linear Model (GLM) to each participant’s fNIRS time series to extract task‐related beta coefficients in six channels for both oxygenated (HbO) and deoxygenated (HbR) hemoglobin.  Second, we computed Inter‐Subject Correlation (ISC) across true versus control pairs to evaluate neural synchrony.  All inferential tests were adjusted for six comparisons (Bonferroni $\alpha=0.05/6$), and we report uncorrected and corrected $p$‐values, effect sizes (Cohen’s $d$), and post‐hoc power.

\subsection*{GLM Results for Hemodynamic Activation}

GLM beta coefficients quantify the magnitude of task‐evoked changes in hemoglobin concentration.  We compared these betas between dual‐player and single‐player conditions using the Wilcoxon signed‐rank test (within‐subject), then applied Bonferroni correction across channels.  Table~\ref{tab:glm_results} presents the full set of statistics, ordered by channel.

Across channels, HbO signals generally exhibited larger, more reliable task effects than HbR.  Channel 1 demonstrated the strongest HbO response (uncorrected $p=0.0057$, $p_{\mathrm{Bonf}}=0.0343$, $d=0.40$, power=0.75), indicating a medium‐sized, well‐powered increase in oxygenated hemoglobin during gameplay.  HbO Channel 2 showed a moderate trend ($p=0.0689$, $d=0.35$) but did not survive correction.  All other HbO channels yielded non‐significant corrected $p$‐values and small to modest effect sizes ($d<0.20$).

Deoxygenated signals (HbR) followed a similar spatial pattern, with Channel 1 again yielding the lowest uncorrected $p$ ($p=0.0252$, $d=0.34$) but failing correction ($p_{\mathrm{Bonf}}=0.1510$).  Remaining HbR channels showed minimal task‐evoked change (all $p_{\mathrm{Bonf}}=1.00$, $d<0.30$) and limited power (<0.40).  These results underscore the greater sensitivity of HbO to block‐design activation in fNIRS.

\begin{table}[ht]
\centering
\caption{GLM beta‐coefficient comparisons (dual vs.\ single) for HbO and HbR across six channels.}
\label{tab:glm_results}
\begin{tabular}{c
                cc cc
                cc cc}
\toprule

  & \multicolumn{4}{c}{HbO} 
  & \multicolumn{4}{c}{HbR} \\
\cmidrule(r){2-5}\cmidrule(l){6-9}
  & $p$     & $p_{\mathrm{Bonf}}$ & $d$    & Power 
  & $p$     & $p_{\mathrm{Bonf}}$ & $d$    & Power \\
\midrule
Ch1 & 0.0017 & 0.0070 & 0.4712 & 0.7493 & 0.0052 & 0.0230 & 0.4292 & 0.5548 \\
Ch2 & 0.0689 & 0.4133 & 0.3514 & 0.6351 & 0.3294 & 1.0000 & 0.0051 & 0.0501 \\
Ch3 & 0.4737 & 1.0000 & 0.1473 & 0.1620 & 0.0884 & 0.5305 & 0.2606 & 0.4014 \\
Ch4 & 0.4532 & 1.0000 & 0.1939 & 0.2464 & 0.6791 & 1.0000 & 0.1379 & 0.1478 \\
Ch5 & 0.3883 & 1.0000 & 0.1850 & 0.2286 & 0.1498 & 0.8987 & 0.2231 & 0.3103 \\
Ch6 & 0.3581 & 1.0000 & 0.0485 & 0.0617 & 0.6306 & 1.0000 & 0.0418 & 0.0587 \\
\bottomrule
\end{tabular}
\end{table}
\subsection*{ISC Results for Neural Synchrony}

To quantify inter‐brain coupling, we computed Inter‐Subject Correlation (ISC) for oxygenated (HbO) and deoxygenated (HbR) hemoglobin signals in six prefrontal channels. ISC was calculated in ten non‐overlapping 90-second windows per session and then averaged across blocks for each channel and pair. True interacting pairs were compared against independent single-player control pairs using the Mann–Whitney U test. To control for multiple comparisons, Bonferroni correction was applied (\(\alpha = 0.05/6\)). We also report Cohen’s \(d\) (using pooled standard deviation for unequal group sizes) and post-hoc power based on observed effect sizes and sample sizes.

Results are presented in Table~\ref{tab:isc_results}. For deoxygenated signals (HbR), Channels 2 and 3 showed significantly higher ISC in true pairs after correction (\(p_{\mathrm{Bonf}}=0.0295\) and \(0.0146\), respectively), each with moderate to large effect sizes (\(d=0.39\) and \(0.64\)) and power above 0.42 and 0.83. Channel 1 exhibited a nominal effect (\(p=0.0236\)) but did not survive correction. Channels 4–6 were non-significant. In contrast, oxygenated signals (HbO) did not reach corrected significance in any channel, with the strongest trend in Channel 6 (\(p=0.0588\), \(d=0.31\), power = 0.30).

Overall, HbR synchrony demonstrated the greatest sensitivity to true interpersonal coupling at Channels 2 and 3, whereas HbO synchrony remained low across all channels under our stringent statistical criteria.  

\begin{table}[ht]
\centering
\caption{ISC statistics for deoxygenated (HbR) and oxygenated (HbO) hemoglobin across six channels.}
\label{tab:isc_results}
\begin{tabular}{c
                cccc
                cccc}
\toprule

  & \multicolumn{4}{c}{HbR}
  & \multicolumn{4}{c}{HbO} \\
\cmidrule(r){2-5}\cmidrule(l){6-9}
  & $p$     & $p_{\mathrm{Bonf}}$ & $d$     & Power
  & $p$     & $p_{\mathrm{Bonf}}$ & $d$     & Power \\
\midrule
Ch1 & 0.0236 & 0.1419 & 0.3393 & 0.3435 & 0.7417 & 1.0000 & 0.0414 & 0.0541 \\
Ch2 & 0.0049 & 0.0295 & 0.3850 & 0.4231 & 0.4419 & 1.0000 & 0.0315 & 0.0524 \\
Ch3 & 0.0024 & 0.0146 & 0.6373 & 0.8317 & 0.4038 & 1.0000 & 0.0011 & 0.0500 \\
Ch4 & 0.9475 & 1.0000 & 0.0475 & 0.0555 & 0.4820 & 1.0000 & 0.1287 & 0.0908 \\
Ch5 & 0.2269 & 1.0000 & 0.2405 & 0.1970 & 0.9475 & 1.0000 & 0.0574 & 0.0580 \\
Ch6 & 0.1296 & 0.7773 & 0.3075 & 0.2917 & 0.0588 & 0.3531 & 0.3145 & 0.3028 \\
\bottomrule
\end{tabular}
\end{table}

\subsection*{Frequency-Specific GLM Effects in Channel 1}

Initial GLM analysis across six prefrontal channels (Table~\ref{tab:glm_results}) revealed that Channel 1 showed the only statistically significant task‐related activation in both oxyhemoglobin (HbO) and deoxyhemoglobin (HbR) signals after Bonferroni correction. To further investigate the source of these effects in the frequency domain, we focused our analysis on Channel 1 and decomposed the fNIRS signal into four frequency subbands using discrete wavelet transform (Daubechies 10, levels 6–9). A block‐design GLM was then applied to each frequency component, and paired statistical testing was conducted to compare dual vs.\ single conditions.

Importantly, because only one channel and four decomposition levels were tested, Bonferroni correction was applied using a divisor of 4. The Wilcoxon signed-rank test—nonparametric and well suited to fNIRS data—was used to evaluate significance. Additionally, Cohen’s $d$ and post-hoc statistical power were computed to assess the magnitude and robustness of effects.

As shown in Table~\ref{tab:glm_wavelet_ch1}, statistically significant activation differences were observed at **Level 9**, the lowest frequency band, for both HbO and HbR. These effects exhibited medium effect sizes ($d \approx 0.43$–$0.47$) and sufficient power (>0.59), indicating that slow hemodynamic fluctuations predominantly contribute to the broadband activation difference originally observed in Channel 1.

\begin{table}[ht]
\centering
\caption{GLM beta‐coefficient contrasts (dual vs.\ single) in Channel 1 across Wavelet levels. Bonferroni correction applied across four levels.}
\label{tab:glm_wavelet_ch1}
\begin{tabular}{lcccccccc}
\toprule

  & \multicolumn{4}{c}{HbO} 
  & \multicolumn{4}{c}{HbR} \\
\cmidrule(r){2-5} \cmidrule(l){6-9}
  & $p$ & $p_{\mathrm{Bonf}}$ & $d$ & Power 
  & $p$ & $p_{\mathrm{Bonf}}$ & $d$ & Power \\
\midrule
6 & 0.1355 & 0.5420 & 0.2350 & 0.1871 & 0.4152 & 1.0000 & 0.1271 & 0.0522 \\
7 & 0.0991 & 0.3963 & 0.1378 & 0.0604 & 0.0162 & 0.0647 & 0.2751 & 0.2696 \\
8 & 0.9542 & 1.0000 & 0.0311 & 0.0145 & 0.1460 & 0.5841 & 0.2081 & 0.1418 \\
9 & 0.0021 & 0.0084 & 0.4665 & 0.6857 & 0.0109 & 0.0436 & 0.4289 & 0.5964 \\
\bottomrule
\end{tabular}
\end{table}

\subsection*{Frequency-Specific ISC in Deoxyhemoglobin (HbR)}

To identify which frequency bands most strongly reflect genuine neural coupling, we computed ISC at wavelet levels 6--9 and applied Bonferroni correction over 8 comparisons (2 channels $\times$ 4 levels). Table~\ref{tab:isc_wavelet_corrected8} presents these results alongside a comparison to the broadband (0.01--0.09~Hz) analysis, where only Channels 2 and 3 achieved significance after correction over 6 channels ($p_{\mathrm{Bonf}} = 0.0295$ and $0.0146$, respectively).

\begin{table}[ht]
\centering
\caption{Wavelet-level ISC statistics for HbR in Channels 2 and 3, Bonferroni-corrected over 8 comparisons.}
\label{tab:isc_wavelet_corrected8}
\begin{tabular}{c|cccc|cccc}
\toprule

  & \multicolumn{4}{c|}{Channel 2} 
  & \multicolumn{4}{c}{Channel 3} \\
\cmidrule(r){2-5}\cmidrule(l){6-9}
  & $p$ & $p_{\mathrm{Bonf}}$ & $d$ & Power & $p$ & $p_{\mathrm{Bonf}}$ & $d$ & Power \\
\midrule
6 & 0.00022 & 0.00179 & 0.43 & 0.51 & 0.80904 & 1.00000 & 0.02 & 0.05 \\
7 & 0.11880 & 0.95040 & 0.31 & 0.29 & 0.40381 & 1.00000 & 0.34 & 0.35 \\
8 & 0.01016 & 0.08128 & 0.32 & 0.29 & 0.61380 & 1.00000 & 0.05 & 0.06 \\
9 & 0.16253 & 1.00000 & 0.25 & 0.20 & 0.00173 & 0.01384 & 0.89 & 0.98 \\
\bottomrule
\end{tabular}
\end{table}

The wavelet-based decomposition reveals a striking pattern of frequency-specific neural coupling that significantly refines our understanding beyond broadband analysis. At Level 6 (0.078--0.157~Hz), Channel 2 demonstrates highly significant ISC enhancement ($p_{\mathrm{Bonf}} = 0.0018$, $d = 0.43$, power $\approx 0.51$), indicating that higher-frequency components carry substantially stronger inter-brain coupling than captured in the original broadband analysis. Conversely, Channel 3 remains non-significant at this higher frequency range, suggesting distinct neural mechanisms underlying synchronization across cortical sites. Moving to intermediate frequencies, Level 7 (0.039--0.087~Hz) shows moderate effect sizes for both channels but insufficient statistical power to survive multiple comparison correction, while Level 8 (0.0195--0.0392~Hz) reveals a marginal trend in Channel 2 ($p_{\mathrm{Bonf}} = 0.0813$, $d = 0.32$) that approaches significance without reaching the corrected threshold. The most compelling finding emerges at Level 9 (0.0098--0.0196~Hz), where Channel 3 exhibits robust ISC enhancement ($p_{\mathrm{Bonf}} = 0.0138$, $d = 0.89$, power $\approx 0.98$) that matches and slightly exceeds its broadband performance, while Channel 2 shows no meaningful effect at this lowest frequency band. These complementary results establish that genuine neural synchrony manifests through distinct frequency-specific mechanisms: Channel 2 couples primarily through higher-frequency oscillations (Level 6), whereas Channel 3 synchronizes via lower-frequency rhythms (Level 9). This frequency-domain dissociation not only demonstrates greater statistical precision and effect sizes compared to broadband ISC but also provides crucial insights into the neural mechanisms underlying inter-brain coupling, emphasizing the fundamental importance of frequency-resolved analysis for understanding how different cortical regions contribute to shared neural states during social interaction.

\subsection*{Supplementary LSTM Classification Results}

To examine whether any consistent temporal pattern existed due to the fixed order of gameplay conditions, we trained a Long Short-Term Memory (LSTM) neural network on windowed fNIRS time-series data. Each input was a 60-second signal, and for each participant, 10 such segments were extracted per condition, resulting in 450 samples for the single-player mode and 450 for the two-player mode. 

The model’s performance, evaluated using 5-fold cross-validation, is reported in Table~\ref{tab:lstm_results}. With an average accuracy of 54.02\% and an AUC of approximately 0.58, the model showed limited discriminative ability. These results suggest that the LSTM was unable to detect a consistent temporal pattern that would differentiate the two experimental conditions. This supports the conclusion that time-dependent confounds such as fatigue or learning effects did not systematically bias participants' neural activity.

\begin{table}[h!]
\centering
\begin{tabular}{lcccccc}
\toprule
Metric     & Accuracy & Loss   & F1-Score & Precision & Recall & AUC \\
\midrule
Mean       & 0.5402   & 0.6891 & 0.5647   & 0.5410    & 0.5925 & 0.5812 \\
Std. Dev.  & 0.0398   & 0.0185 & 0.0362   & 0.0349    & 0.0417 & 0.0473 \\
\bottomrule
\end{tabular}
\caption{Cross-validation performance metrics of the LSTM classifier distinguishing single-player from two-player fNIRS segments.}
\label{tab:lstm_results}
\end{table}

\section*{Discussion}

This study provides novel insights into the neural dynamics of social interaction by contrasting brain activity during engagement with intentional versus artificial agents in a realistic tennis simulation. Through the integration of multiple analytical strategies—General Linear Modeling (GLM), inter-subject correlation (ISC), wavelet-based frequency decomposition, and control machine learning analysis—we found converging evidence that intentional agents elicit distinct and detectable brain responses.

\begin{figure}[!t]
\centering
\hfill
\subfloat[X-Coordination]{\includegraphics[width=0.48\textwidth, height = 0.22\textheight]{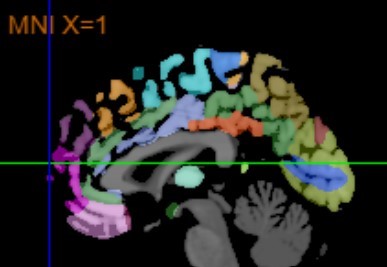}}
\hfill
\subfloat[Y-Coordination]{\includegraphics[width=0.48\textwidth, height = 0.22\textheight]{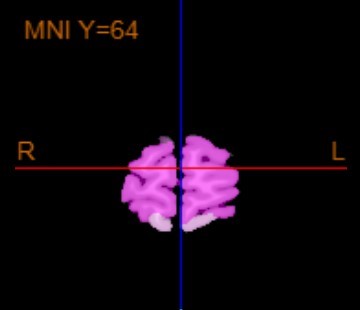}} \\
\centering
\subfloat[Z-Coordination]{\includegraphics[width=0.48\textwidth, height = 0.22\textheight]{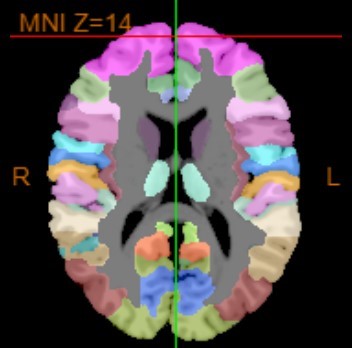}}
\caption{An approximate graphical representation of the channel 1. This is nearly aligned with previous research that identified Brodmann area 10 and the frontal cortex as responsible for social behaviour}
\label{fig:discussion}
\end{figure}

At the activation level, GLM analysis revealed that Channel 1, located in the fronto-medial prefrontal cortex, showed a statistically significant increase in task-evoked HbO signal during interaction with human agents. The effect was especially prominent in the lowest frequency band (0.0098–0.0196 Hz), suggesting the involvement of slow, sustained hemodynamic processes in representing social presence. This pattern supports the view that low-frequency oscillations reflect stable and continuous cognitive states, such as theory of mind or social attribution.
The findings in this analysis are consistent with prior observations using more spatially precise neuroimaging techniques like fMRI and PET \cite{krach2010rewarding, gallagher2002imaging, abu2020re}, affirming the robustness of the results across methodologies. The BioImage Suite website \footnote{\url{https://bioimagesuiteweb.github.io/webapp/}} is used to generate an approximate graphical representation of the brain region illustrating in Fig.\ref{fig:discussion}.

In terms of neural coupling, ISC analysis revealed significantly higher synchrony in true interacting pairs compared to false or non-interacting pairs, especially in Channels 2 and 3 for deoxygenated hemoglobin. Frequency-resolved ISC indicated that different channels operated in different bands: Channel 2 showed coupling in higher cognitive frequencies (e.g., 0.078–0.157 Hz), while Channel 3 synchronized via low-frequency activity (e.g., 0.0098–0.0196 Hz). This frequency dissociation provides strong evidence that multiple cognitive subprocesses—such as prediction, attention, and mental state alignment—are encoded in distinct oscillatory regimes.

To ensure that the observed effects were not artifacts of experimental ordering (i.e., always starting with the single-player condition), we trained an LSTM model on 60-second segments of fNIRS time series to discriminate between task conditions. The model achieved low classification performance (accuracy $\approx$ 54\%, AUC $\approx$ 0.58), indicating that temporal trends such as fatigue or task familiarity did not substantially bias the results. This supports the validity of our main findings and reinforces that brain activity differences are primarily driven by the nature of the opponent.

Notably, the ISC findings provide promising groundwork for future applications of deep learning in real-world cognitive modeling. With more channels, larger datasets, and richer interaction settings, it may become feasible to construct models capable of identifying fine-grained social dynamics. For example, in our simulated tennis game, one could potentially develop a neural decoding system that determines \emph{who} a participant is playing against solely from their fNIRS signals—an exciting prospect for real-time brain-based interaction analysis.

In summary, our study demonstrates that social cognition—specifically the perception of intentionality—is encoded both in localized brain activity and in interpersonal neural synchrony. These findings not only advance the neurocognitive understanding of human-agent interaction but also highlight the practical potential of fNIRS and machine learning for decoding social behavior in naturalistic settings. Future studies with broader populations and extended sensor coverage could further enhance the resolution and applicability of these methods.

\section*{Conclusion}
This study leveraged fNIRS hyperscanning during a dynamic, game-based tennis paradigm to compare neural responses when interacting with an intentional human agent versus an artificial intelligence opponent. General Linear Model analysis revealed significantly greater task-evoked activation in the medial prefrontal cortex (Channel 1) during human–human play, driven primarily by slow hemodynamic oscillations (0.0098–0.0196Hz). Intersubject Correlation showed enhanced deoxyhemoglobin synchrony in true dyads at Channels 2 and 3, with frequency-specific coupling: higher-frequency alignment in Channel 2 (0.078–0.157Hz) and low-frequency alignment in Channel 3 (0.0098–0.0196Hz). By contrast, AI interactions elicited reduced cortical engagement and no measurable inter-brain synchrony. A control LSTM analysis ruled out systematic order effects, confirming that observed differences reflect perceived agency rather than fatigue or learning. Together, these findings provide robust neural evidence that the perception of intentionality modulates both localized activation and interpersonal synchrony, underscoring the utility of fNIRS hyperscanning for studying naturalistic social cognition. The results have direct implications for the design of socially interactive AI systems that aim to engage human cognitive and affective processes more effectively.

\section*{Limitations}
\begin{itemize}
  \item \textbf{Sample demographics:} All participants were right-handed males aged 18–30, limiting generalizability across genders, handedness, and age groups.
  \item \textbf{Spatial coverage:} fNIRS optodes targeted prefrontal regions only, precluding insight into other nodes of the social brain network (e.g., temporoparietal junction, superior temporal sulcus).
  \item \textbf{Fixed task order:} The solo (AI) condition always preceded the human–human condition; although LSTM control analysis showed no major order effects, a fully counterbalanced design would strengthen causal inference.
  \item \textbf{AI opponent realism:} The computer agent’s behavior, while algorithmically driven, did not vary in apparent intentionality; future work should compare different AI “personalities” or adaptive strategies.
  \item \textbf{Physiological confounds:} Although both oxy- and deoxyhemoglobin signals were analyzed, residual systemic artifacts (e.g., heart rate, respiration) may have influenced low-frequency components.
  \item \textbf{Ecological validity:} The virtual tennis paradigm, despite its dynamic nature, represents a specific competitive context; extending to cooperative or conversational tasks would broaden applicability.
\end{itemize}

\bibliography{main}







\end{document}